\documentclass[12pt,english]{paper}
\usepackage[T1]{fontenc}
\usepackage[latin9]{inputenc}
\usepackage{babel}
\usepackage{subfigure}
\usepackage{amsmath}
\usepackage{amsfonts}
\usepackage{amsthm}
\usepackage{mathtools}
\usepackage{amssymb}
\usepackage{dsfont}
\usepackage{float}
\usepackage{setspace}
\usepackage{bbm}
\usepackage{rotating}

\usepackage{color} 

\usepackage{graphicx}
\usepackage[normalem]{ulem}
\usepackage{geometry}
\usepackage{natbib}

\usepackage{hyperref}
\hypersetup{
    colorlinks=false,
    linkcolor=blue,
    filecolor=magenta,      
    urlcolor=cyan,
}
 
\usepackage{xargs} 
\usepackage[pdftex,dvipsnames]{xcolor}  

\usepackage[disable]{todonotes} 
\newcommandx{\andrea}[2][1=]{\todo[inline,linecolor=red,backgroundcolor=red!25,bordercolor=red,#1]{\textbf{Andrea}: #2}}
\newcommandx{\vincent}[2][1=]{\todo[inline,linecolor=blue,backgroundcolor=blue!25,bordercolor=blue,#1]{\textbf{Vincent}: #2}}

\usepackage{tikz}
\usepackage{pgfplots}
\usetikzlibrary{calc,matrix}

\makeatletter
\theoremstyle{plain}

\theoremstyle{plain}
\newtheorem{lemma}{\protect\lemmaname}
\theoremstyle{plain}
\newtheorem{proposition}{\protect\propositionname}
\theoremstyle{definition}

\theoremstyle{plain}

\theoremstyle{plain}

\theoremstyle{plain}

\makeatother

\providecommand{\definitionname}{Definition}
\providecommand{\lemmaname}{Lemma}
\providecommand{\propositionname}{Proposition}
\providecommand{\theoremname}{Theorem}
\providecommand{\assumptionname}{Assumption}
\providecommand{\corollaryname}{Corollary}
\providecommand{\remarkname}{Remark}

\begin{document}

\title{ Commitment Against Front-Running Attacks\thanks{We are grateful to the editor Joshua Gans, two anonymous referees, Agostino Capponi, Jiasun Li, Christof Ferreira Torres, Arthur Gervais, Ari Juels, and the participants to UBRI Connect 2022, Tokenomics 2022, ETHcc 2023 for their comments and suggestions. We gratefully acknowledge the financial support of the Ethereum Foundation (grant FY22-0840).}}

\author{Andrea Canidio
\thanks{CoW Protocol}  and Vincent Danos\thanks{CNRS and  \'{E}cole Normale Sup\'{e}rieure (France); }
}

\maketitle

\vspace*{-.5 cm}

\noindent \today

\begin{abstract} 
We provide a game-theoretic analysis of the problem of front-running attacks. We use it to distinguish attacks from legitimate competition among honest users for having their transactions included earlier in the block. We also use it to introduce an intuitive notion of the severity of front-running attacks. We then study a simple commit-reveal protocol and discuss its properties. This protocol has costs because it requires two messages and imposes a delay. However, we show that it prevents the most severe front-running attacks while preserving legitimate competition between users, guaranteeing that the earliest transaction in a block belongs to the honest user who values it the most. When the protocol does not fully eliminate attacks, it nonetheless benefits honest users because it reduces competition among attackers (and overall expenditure by attackers). \\
\textbf{Keywords}: Front running, Game theory, MEV, Transactions reordering, commit-reveal
\end{abstract}

\section{Introduction}
On the Ethereum network, each validator decides how to order pending transactions to form the next block, hence determining the order in which these transactions are executed. As a consequence, users often compete with each other to have their transactions included earlier in a block, either by paying transaction fees or by making side payments directly to validators.\footnote{Competition through higher transaction fees occurs via ``gas replacement'' transactions, whereby a pending transaction is resubmitted with a higher fee. The resulting game is akin to an auction (see \cite{daian2019flash}). The most common way to make side payments to validators is to use flashbots (see \url{https://github.com/flashbots/pm}).} This form of competition can be beneficial because it ensures that a scarce resource (i.e., having a transaction included earlier in the block) is allocated to the user who values it the most.\footnote{Whether it is the most efficient to achieve this goal is a different issue we do not address here.} But at the same time, it opens the possibility of front-running attacks: because pending transactions are public, a malicious user can observe a victim's incoming transaction, craft a new transaction and then pay to place it before that of the victim.

Importantly,  legitimate competition and  attacks are often difficult to distinguish.   As an illustrative example, consider a smart contract programmed to award a valuable NFT to the first person who correctly answers a question. Assume that the smart contract does not have an explicit mechanism to resolve competing claims to the object (i.e., by running an auction among those who provided the correct answer) and settles claims in order of arrival. In this example, competition between users can arise in two cases. In the first case, two users simultaneously and independently find the answer. Each submits it and competes to have his/her transaction included earlier in the block. Because the user who values the NFT the most is willing to pay more, this user should be able to place his transaction before that of the opponent, thereby winning the NFT. In the second case, an honest user finds the answer and sends it to the smart contract. A malicious user observes the transaction, copies it, and competes to have its copy included in the block earlier than the original transaction. 

From the observational point of view, the above two situations are identical: two users submit the same answer and then compete to have it included earlier in the block. Despite this, the first is an example of legitimate competition because  each user would have submitted his answer also in the absence of the other user. The second is an attack because the attacker cannot send his transaction if he does not observe the victim's transaction. 
Note also that  the extent to which an attacker relies on the victim's transaction can be interpreted as a measure of the \textit{severity} of a front-running attack.  For example, the ``attacker'' could be another user who, through his research, narrowed the correct answer to two or three possibilities.     This attack seems less severe than one in which the attacker has no prior information.

In this paper, we propose a game-theoretic model of front-running. Our goals are two. First, inspired by the above discussion, we aim to provide a  formal definition of front-running attacks (vs.\@ legitimate competition among honest users) and their severity. Second, we use the model to study a simple commit-reveal protocol inspired by the ``submarine commitment'' protocol  (\citealp{breidenbach2018enter}) and by the ENS domain registration, which we discuss later.  The protocol we study can be implemented at the smart contract level without modifying neither the underlying Ethereum infrastructure, nor introducing third parties or layer-2 networks. 
In its simplest version, the user concatenates the desired message  with the address from which the reveal message will be sent and passes this into a function with an intractable pre-image problem (for example,  the SHA-256 hash function). The resulting output is the commit message, which the user sends to the smart contract. 
Then, the user sends the reveal message to the smart contract, where the reveal message is simply the desired message. The smart contract receiving a reveal message will execute it only if the concatenation of the reveal message with the address from which it was received corresponds to the commit message. 

Our model highlights several benefits of the commit reveal protocol. The key observation is that an attack involves two steps: (i)  committing a message without knowing what message the victim will send and (ii) after observing the victim's reveal message,  deciding to send the committed message or no message at all. 
The protocol, therefore, forces the attacker to make a costly guess: send a costly commit message without knowing what the victim committed. Hence,  the protocol can eliminate front-running attacks, especially when it is difficult for an attacker to guess. 
By definition of the severity of the attack introduced earlier, we can say that our protocol is most effective when the severity of the attack is high and is less effective when the severity of the attack is low. At the same time, the protocol does not impede legitimate competition between users: two honest users can commit their messages and then compete to have their reveal message included earlier in the block.

We then study the case in which multiple potential attackers are present (see Section \ref{sec: multiple attackers}). Absent the protocol, competition  pushes each attacker to overspend  (relative to the single-attacker case). This is detrimental to both the attackers and the honest user. In particular, all attackers but the strongest earn zero in expected terms.\footnote{This result is a version of a well-known result in contest theory, that of ``full dissipation of rents''. See, for example, \cite{fudenberg1987understanding}.	}  Instead, in the commit-reveal protocol, the commit message acts as a fixed cost to  attack in the next period. Because in the following period, the weakest attackers earn zero, there is no equilibrium in which all attackers send the commit message with probability 1: either there is no attack; or a single attacker commits and then attacks; or the attackers commit (and then attack) with probability strictly less than one. As a consequence, in the best-case scenario, the protocol eliminates attacks; in the worst-case scenario, it reduces the level of competition between attackers resulting in fewer resources spent in an attack, which benefits the honest player as well.


As an extension (see Section \ref{sec:hiding commitment}), we study a variation of the above protocol in which the identities of the sender and of the receiver of the commit message are obfuscated. This variation hinges on the existence of a template code for a container smart contract. When committing, the user uses a brand new address  to create a container smart contract using the template and then sends the commit message to this newly-created container, which time-stamps the commitment message with the current block number. When sending the reveal message, the honest user also sends a pointer to the container smart contract where the commitment is located. The smart contract considers the commitment valid if the commit message is correct, its timestamp is antecedent to the current block, and, crucially, if the code of the container smart contract corresponds to the template. This way, an outside observer can only see that someone created a commitment smart contract and sent a commit message, but not who committed nor the target smart contract for that commitment. Guessing is even harder for an attacker, and hence the probability of an attack is even lower.

\paragraph{Prior work}

Our commit-reveal protocol is novel but similar to existing proposals. Our main contribution is, therefore, not  the protocol itself but its theoretical analysis: we use a game-theoretic model to show that such protocol (i) eliminates the most severe front-running attacks,  (ii) maintains legitimate competition, and (iii) reduces competition among attackers. Also, most of the literature has studied solutions to reduce or eliminate front-running by  changing Ethereum's infrastructure or introducing third parties (See  \cite{heimbach2022sok} for a literature review). We instead focus on a solution that does not require third parties and can be implemented at the smart contract level, allowing for flexibility in its implementation. For example, each smart contract could decide that only some messages must follow the protocol to be considered  valid, while others do not need to (see  Section \ref{sec: partial implementation}). Or a smart contract may decide that the protocol is required only during some   periods (see Section \ref{sec: waiting in the mempool}). Finally, as we discuss later, existing solutions are primarily concerned with eliminating attacks (at the cost of eliminating legitimate competition) \textit{or} better organizing competition (at the cost of exacerbating attacks). Our results show that commit reveal protocols can eliminate attacks while maintaining legitimate competition.

Our protocol can be seen as a simplified version of the submarine commitments protocol in \cite{breidenbach2018enter}: in both cases, a message is first committed and then revealed, and the commitment can be hidden in the sense that the identity of the sender and receiver of the commit message cannot be observed. The main difference is that we adopt a weaker notion of  ``commitment'' because we allow users not to send a transaction after committing it. The notion of ``commitment'' in \cite{breidenbach2018enter} is instead stronger because users are penalized for not following through with their commitment. 

Our protocol is similar to the commit-reveal scheme used by the Ethereum Name Service (ENS), the service managing the domain registration on the Ethereum network, that is, the association of a human-readable name to an Ethereum address. The problem of front running arises because registering a domain may reveal that it is valuable (perhaps because it corresponds to the name of a soon-to-be-launched service). Hence, an attacker could purchase a domain by front-running an honest user and then try to re-sell the same domain to the same honest user. For this reason, to register a domain with ENS, a user must first send a commit message, which  is the hash of the  domain name and the address that will own it.\footnote{See the documentation \url{https://docs.ens.domains/contract-api-reference/.eth-permanent-registrar/controller} (accessed on July 28th, 2023). The fact that the address that will own the domain is part of the commit message is not discussed in the text, but appears in the sample code provided (see the variable ``owner''), and was confirmed to us orally by representatives of ENS.} Our scheme generalizes the ENS approach by replacing the address that will own the domain with the address from which the revealed message will be sent (which, in general, uniquely identifies the participants to the blockchain).



As already mentioned,  we provide a game-theoretic analysis of the properties of this protocol,  applicable to any smart contract.\footnote{\cite{breidenbach2018enter} analyze the properties of the submarine commitment scheme in the context of a bug-bounty scheme they propose.} In this regard, our work draws inspiration from~\cite{gans2022solomonic}, who  develop a game-theoretic analysis of the problem of front-running arising when an honest user and an attacker claim the same reward (see also the discussion in \cite{Gans2023}). They also propose a protocol that eliminates these types of attacks.  
 Their key assumption is that the legitimate claimant strictly prefers the reward to be burned rather than paid to the attacker, and hence it can be applied to smart contracts that can destroy resources that should be allocated to their users.

 

Flashbots is a well-known project aiming to better organize competition among users. The premise  is that competition through transaction fees  can lead to so-called ``gas wars'' by which a given block is filled with transactions that will fail (because only the first one can be correctly executed). Gas wars impose a negative externality on all users because they lead to congestion  and higher transaction fees. The idea is to eliminate these negative externalities by allowing users to pay validators directly, therefore keeping their messages private.\footnote{ We note that our protocol also reduces competition among attackers and hence reduces "gas wars".}  Doing so, however, makes it extremely easy to attack an honest user who sends his or her message publicly (see \cite{https://doi.org/10.48550/arxiv.2202.05779}).

Other proposed solutions either attempt at encrypting all transactions, or at imposing exogenous criteria for ordering transactions, or both, therefore preventing attacks but also hindering legitimate competition. \cite{cryptoeprint:2020/269} propose the Aequitas protocol, a method to achieve ordering-fairness, by which if sufficiently many nodes observe a given transaction arriving before a different transaction, this ordering should be maintained when these transactions are included in a block.\footnote{See also the hedera-hashgraph project (\cite{baird2016swirlds}).}   There are also commit-reveal schemes intermediated by  third parties in charge of, for example, reorganizing incoming transactions  while also encrypting and then decrypting them. In this context, a widely recognized solution is the Shutter network, in which a network of nodes called ``keypers'' jointly generate cryptographic keys with which users encrypt their transactions.   Users then submit these transactions to a batcher contract that also orders them. Finally, Keypers broadcast the decryption key, the transactions are decrypted and sent to the target smart contracts.

A concept that is often associated with front-running attacks is that of Maximal-extractable value (MEV), defined as ``the maximum value that can be extracted from block production in excess of the standard block reward and gas fees by including, excluding, and changing the order of transactions in a block'.'\footnote{See \url{https://ethereum.org/en/developers/docs/mev/}.} Most existing measures of total MEV are not very useful in our context as they capture both users' legitimate competition (sometimes called ``good MEV'') and attacks (sometimes called ``bad MEV''). A few papers, however, identify specifically profits extracted from attacks. \cite{torres2021frontrunner} collect on-chain data from the inception of Ethereum (July 30, 2015) until November 21, 2020. They estimate that these attacks generated 18.41M USD in profits for the attackers, of which 13.9M USD due to sandwich (also called \textit{insertion}) attacks.\footnote{A sandwich or insertion attack occurs when an attacker simultaneously front-runs and back-runs a victim's transaction. They are common in the context of Automated Market Makers (AMM),  the dominant type of decentralized exchange.  When a user wants to perform a swap, an attacker will front-run the victim with the same swap and then back-run her with the opposite swap. Doing so allows the attacker to ``buy cheap'' and ``sell expensive'' while forcing the victim to trade at less favorable terms.} They also identify instances where several attackers competed to attack the same victim. Similarly, \cite{qin2022quantifying}  consider a later period (from the 1st of December, 2018 to the 5th of August, 2021) and find that sandwich attacks generated  174.34M USD in profits.

\andrea{another useful pointer to existing approaches/solutions to front running attacks https://www.mev.wiki/solutions}

\andrea{According to the guy from the Ethereum foundation, we  should discuss this paper: \url{https://arxiv.org/pdf/2109.04347.pdf}  Apparently they also have some definition of what a front-running attack is. But, after skimming through it, I don't think it is related to what we do.}

\section{The problem: front-running attacks}

We start by developing a model of front-running attacks and later introduce the commit-reveal protocol. 
There is a smart contract $SC$ and two players: $A$lice and $B$ob. There is a piece of information, call it ``the state of the world,'' $s\in S $ that only $A$ learns at the beginning of the game. Absent front running attacks, after observing $s$, player $A$ sends a message $\tilde \sigma_A \in \Sigma $ to the mempool (i.e., the set of pending transactions), where $\Sigma$ is the space of possible messages. Both the set $S$ and the set $\Sigma$ are non-empty and finite. When the message $\tilde \sigma_A$ is included in a block, the smart contract $SC$ performs an action that generates a benefit $\tilde P_A(\tilde \sigma_A, s)$ to player $A$. 

Front-running attacks arise because messages in the mempool are public. Hence, after $A$ sends a message to the mempool, this message is observed by $B$, who can send a  counter-message $\tilde \sigma_B \in \Sigma $. If $\tilde \sigma_B $ is included in the blockchain before A's message, then $B$ earns  $\tilde P_B(\tilde \sigma_B, \tilde \sigma_A, s)$ while $A$ earns nothing. Else, $B$ earns nothing and $A$ earns $\tilde P_A(\tilde \sigma_A, s)$. 

 Sending messages is costly. Each player can send a regular message by paying $c>0$. If multiple regular messages are sent, they are included in the block in the order they are sent. We can think of $c$ as being the base fee: a fee that should guarantee the  inclusion of a transaction in the next bloc, at least outside of periods of rapid change in the demand for transactions.\footnote{\label{footnote: EIP1559}The concept of base fee was introduced with the  EIP-1559 upgrade. See the original proposal here \url{https://eips.ethereum.org/EIPS/eip-1559}. For an economic analysis of EIP-1559, see \cite{roughgarden2020transaction}.} 
 Player $B$, however, can also pay $f>c$ to send a ``fast'' message that, with probability $q$,  is included in the block before  $A$'s regular message, despite $A$'s message being sent first. For example, $f$ could be the cost of sending a transaction via a service such as flashbots, or could be a regular mempool transaction with a transaction fee significantly above the base fee.   Here we consider the parameters $q$, $c$, and $f$ as exogenous and determined by the technology available to $A$ and $B$. We relax this assumption in Section \ref{sec: multiple attackers}, in which we introduce multiple $B$ players choosing their  own $f$, which then determine the probability that a given $B$ player successfully front runs both $A$ and the other $B$ players. 

Regarding applications, consider the example we discussed in the introduction: a  smart contract that rewards whoever can correctly answer a question. In this case, $B$ will learn the correct answer by observing $A$'s message and then try to submit the same answer before $A$. Formally, $s= \sigma_A(s)=\sigma_B(s)$. Our model also fits a famous (nonfictional) example discussed in the blog post ``Ethereum is a dark forest'' (\citealp{robinson_konstantopoulos_2020}). In this example, two researchers wanted to recover some tokens previously sent to an incorrect address. They realized that anyone who knew about their existence could have recovered them. Despite their effort, they were front-run by an attacker who managed to steal them. In the context of our model, again $\sigma_A(s)=\sigma_B(s)$. Another fitting example is that of  an AMM. Player $A$ is a liquidity provider who, upon learning some private information $s$, decides to withdraw some or all the liquidity provided. By observing such a message, $B$ can infer that something has changed in the environment and try to steal the same liquidity. In this case,  $\sigma_A(s)=$\{withdraw my liquidity\}, $\sigma_B(s)=$\{swap some tokens\}.\footnote{For a study of this type of attack, see \cite{capponi2021adoption}. For a  study of similar attacks in the context of traditional exchanges, see Section 6 of \cite{budish2015high}. } Also relevant in the context of AMMs are sandwich attacks, in which $A$ sends message $\sigma_A(s)=$\{swap some tokens\}, and $B$ then front runs $A$ with a message $\sigma_B(s)=$\{perform the same swap as $A$\} and ``back-runs'' $A$ with the message $\sigma_B(s)=$\{perform the opposite swap as $A$\}. This attack is profitable because it exploits the slippage curve of the AMM. Although we do not explicitly allow $B$ to back-run $A$, our analysis extends to sandwich attacks because front-running is a necessary component of a sandwich attack (modulo the fact that sandwich attacks are more costly than simple front-running attacks because they require an additional message).

We make two simplifying assumptions. First, we assume that $A$ is partially naive. She is naive in that she always chooses the message that maximizes her payoff given the state of the world; however, she is sophisticated in whether to send her message (or, in the next section, to initiate the protocol). Therefore, we rule out the possibility that $A$ chooses her message to manipulate $B$'s belief about the state of the world, which we think is unrealistic.\footnote{This assumption does not play any role if front running attacks do not occur in equilibrium (for example, because of the commit-reveal protocol we introduce in the next section). That is, if there are no attacks when Alice is partially naive, then there are no attacks when Alice is fully sophisticated. If $A$ is fully sophisticated, then under some parameters the equilibrium of the game is a partition of the possible states of the world $S$ such that $A$ sends the same message in  all states of the world belonging to the same part of the partition. Upon observing the message, $B$ learns the part of the partition but not the exact state of the world. The results for a given partition are identical to those presented  here. However, deriving the equilibrium partition is non-trivial and of second-order importance relative  to our main research question.   } Mathematically,  after observing the state of the world, if $A$ sends a message, she sends a message 
\[
\sigma_A(s) \equiv \mbox{argmax}_{\tilde \sigma_A \in \Sigma} \tilde P_A(\tilde \sigma_A, s). 
\]
Given this, we can re-define $A$'s payoff in case she sends a message and she is not front-run as:
\[
P_A(s) \equiv \tilde P_A(\sigma_A(s), s).
\]
The second simplifying assumption is that $\sigma_A(s)$ is a bijection: in each state of the world, there is a unique and distinct message maximizing  player $A$'s payoff. This a useful simplification because $A$'s message (if sent and observed) always reveals the state of the world. It follows that $B$'s optimal counter message after observing $\sigma_A(s) $  and learning $s$ is:
\[
\sigma_B(s) \equiv \mbox{argmax}_{\tilde \sigma_B \in \Sigma} \tilde P_B(\tilde \sigma_B, \sigma_A(s), s) .
\]
The resulting payoff for player $B$ if he successfully front-runs $A$ is:
\[
P_B(s) \equiv \tilde P_B(\sigma_B(s), \sigma_A(s), s).
\]

\paragraph{Equilibrium}

\begin{figure}[h]
\tikzset{
solid node/.style={circle,draw,inner sep=1.5,fill=black},
hollow node/.style={circle,draw,inner sep=1.5}
}
\begin{center}

\begin{tikzpicture}[scale=1.5,font=\footnotesize]
\tikzstyle{level 1}=[level distance=15mm,sibling distance=35mm]
\tikzstyle{level 2}=[level distance=15mm,sibling distance=35mm]
\node(0)[solid node,label=above:{$A$}]{}
child{node(1)[solid node,label=above:{$B$}]{}
child{node[hollow node,label=below:{$((1-q) P_A(s)-c,~q P_B(s) - f)$}]{} edge from parent node[left]{$\sigma_B (s)$}}
child{node[hollow node,label=below:{$(P_A( s)-c,0)$}]{} edge from parent node[right]{no message}}
edge from parent node[left,xshift=-3]{$\sigma_{A}(s)$}
}
child{node(2)[hollow node,label=below:{$(0,0)$}]{}
edge from parent node[right]{no message}
};
\end{tikzpicture}

\end{center}
\caption{Game tree for given $s$.}\label{fig:game tree benchmark}
\end{figure}
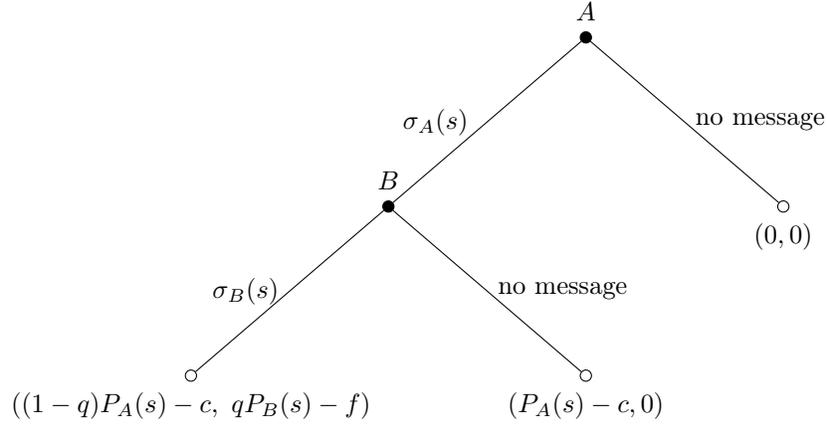

The above assumptions allow us to write the extensive form of the game for given $s$ as in Figure \ref{fig:game tree benchmark}, which we can easily solve by backward induction. If $A$ sends a message, then $B$ attempts to front-run if and only if:
\[
q P_B(s)>f
\]
Given this, we can derive  $A$'s optimal strategy. Suppose the state of the world is such that $q P_B(s)<f$, and $A$ expects no front running. In this case, she sends a message if and only if
\[
P_A(s)>c
\]
If, instead, the state of the world is such that $q P_B(s)>f$, then $A$ anticipates that $B$ will try to front-run. In this case, $A$ sends a message if and only if 
\[
(1-q) P_A(s)>c
\]

The following proposition summarizes these derivations.
\begin{proposition}[Equilibrium]\label{prop:equilibrium}
Player $A$'s equilibrium strategy is:
\begin{equation}
    \sigma^*_A(s) = \begin{cases}  
\emptyset &\mbox{ if } P_A(s)<c \mbox{ or } q P_B(s)>f  \mbox{ and } (1-q)  P_A(s)<c \\
\sigma_A(s) &\mbox{ otherwise }
    \end{cases}
\end{equation}
where $\sigma^*_A(s) =\emptyset$ means that $A$ does not send any message. Player $B$'s equilibrium strategy is
\begin{equation}
    \sigma^*_B(s) =  \begin{cases} \sigma_B(s) &\mbox{if } q P_B(s)>f \mbox{ and }  \sigma^*_A(s) \neq \emptyset   \\ \emptyset &\mbox{otherwise} 
    \end{cases}
\end{equation}
\end{proposition}
Hence, front running does not happen when its benefit is low (i.e., $P_B(s)\leq f/q$). If, instead, its benefit is large (i.e., $P_B(s) > f/q$), $B$ will attempt to front run $A$ whenever $A$ sends a message. In particular, when $P_A(s)>c$ but $(1-q) P_A(s)<c$ the threat of front running prevents $A$ from sending the message in the first place, therefore destroying the value of the exchange between $A$ and $SC$.

\paragraph{Front-running attacks vs.\@ legitimate competition.}

In the introduction, we argued that the difference between front-running attacks and legitimate competition is whether the ``attacker'' relies on the information extracted from observing the victim's  message. This intuitive notion can be formalized by considering a modified game in which player $B$ is under a ``veil of ignorance'': he chooses whether to send his message and what message to send without knowing whether $A$ sent a message or what message she sent.   We want to find necessary and sufficient conditions such that, in the equilibrium of this modified game, $B$ does not want to send any message. Clearly, if $B$ does not send any message, then $A$'s optimal strategy is simply:
\begin{equation}\label{eq: optimal no front running}
\sigma^{**}_A(s) \equiv \begin{cases}   \sigma_A(s) &\mbox{if } P_A(s)\geq c \\ \emptyset &\mbox{otherwise} \end{cases}
\end{equation}
Assuming that $B$ knows the probability distribution over the possible states of the world, there is an equilibrium in which $B$ does not send any message if and only if
\begin{equation}\label{eq: assumption}
q \cdot E_s[\tilde P_B(\tilde \sigma_B, \sigma^{**}_A(s), s)] \leq f ~~~\forall \tilde \sigma_B \in \Sigma. 
\end{equation}

In what follows,  if in the equilibrium of the original game, $B$ sends a message and condition \ref{eq: assumption} holds, then we say that there is a front-running attack.  If instead,  in the equilibrium of the original game, $B$ sends a message and condition \ref{eq: assumption} is violated, then we say that $B$ is a legitimate competitor.\footnote{It is possible that \eqref{eq: assumption} does not hold and hence  $B$ sends a message also when he does not observe $A$'s actions. At the same time, he may choose a different message if he observes $A$'s message. According to our definition, this is not a front-running attack, even if $ B$ uses $A$'s message. This is justified by the observation that, in our model, $A$'s payoff does not depend on what message $B$ sends.} As we will see, this distinction will play an important role in the next section when we introduce our commit-reveal protocol. The reason is that the protocol reduces (but not fully eliminates) $B$'s ability to act upon $A$'s message. If \eqref{eq: assumption} holds, the expected benefit of an attack is reduced, and hence attacks are less likely. If instead \eqref{eq: assumption} is violated, then $B$ always has a profitable message to send, independently of his observation of $A$. In this case, the protocol has little impact on $B$'s behavior, except for requiring him to send two messages. This means that the protocol  reduces the expected return of an attack (i.e. when \ref{eq: assumption} holds) but has little impact on legitimate competition (i.e., when \ref{eq: assumption} is violated)

\section{Preventing front-running via commitment}

To address the problem of front-running attacks, here we propose a commit-reveal protocol. In terms of notation, we call player $A$'s commit message $\sigma_{A,1}$  and reveal  message $\sigma_{A,2}$. Similarly, player $B$'s counter-messages are $\sigma_{B,1}$ and  $\sigma_{B,2}$.

\begin{figure}[ht]
    \centering
    \includegraphics[width=15cm,trim={0 8cm 0 2cm}]{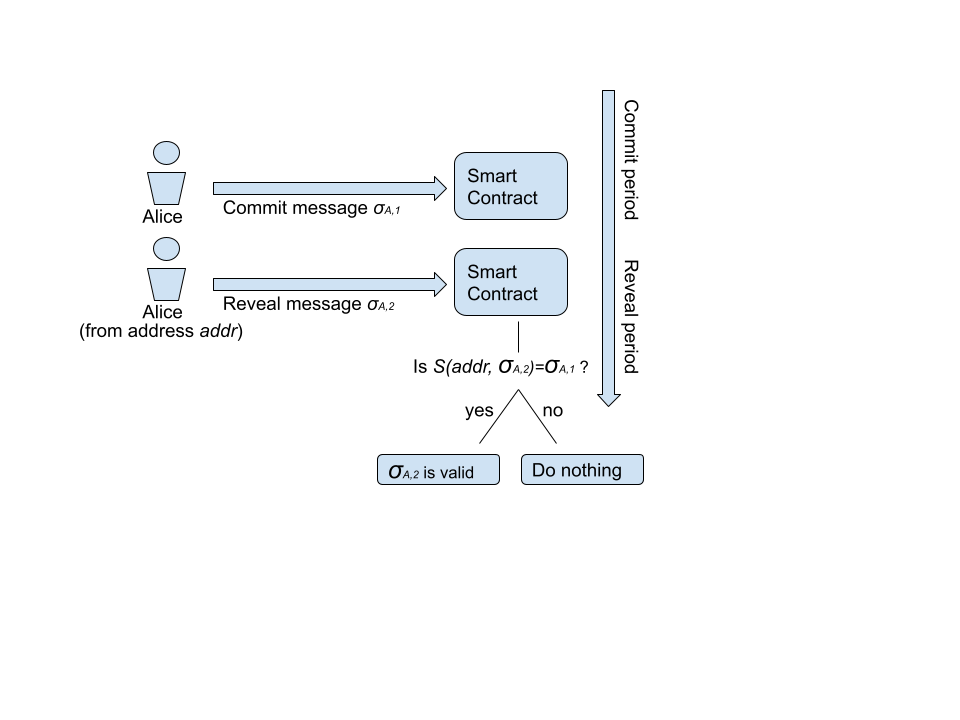}
    \caption{The commit-reveal protocol}
    \label{fig:protocol}
\end{figure}

Formally,  the protocol has a commitment period and a reveal period, which here are   two subsequent blocks.\footnote{In Section \ref{sec: waiting in the mempool} we discuss more in detail the problem of specifying commit and reveal periods. }
If player $A$ wants to send message $\sigma_A \in \Sigma $ to $SC$, in the commit period  $A$ sends the commit message 
\[\sigma_{A,1} = S(addr,\sigma_A)\]
 to $SC$ where $addr$ is an address  that  $A$ controls and  $S()$ is a function with an intractable pre-image problem (for example, $Hash\left(addr| \sigma_A   \right)$ where $Hash()$ is the SHA-256 hash function).  Once the commit  message is included in a block,     $A$  sends the reveal message $\sigma_{A,2}=\sigma_A $ to $SC$ from the address $addr$, which is then included in the next block. Upon receiving this message, $SC$ computes $S(addr,\sigma_A)$ and checks it against the previously received commit messages. If there is a commit message that is equal to $ S(addr,\sigma_A)$, then SC performs a given action generating a payoff to player $A$. Else, nothing happens. See Figure \ref{fig:protocol} for an illustration.

Hence, if multiple commit messages and reveal messages are sent, the smart contract will execute the first valid reveal message included in a block. It follows that player $B$ can attack by sending a commit message (as a regular transaction) and then a reveal message as a fast transaction. The attack is successful if $B$'s reveal message is included in a block before $A$'s reveal message.

There is a common discount factor $\beta\in[0,1]$, so when a given payoff is earned with a block delay, this payoff is discounted by $\beta$. We assume that  $A$ does not observe $B$'s commit message and hence cannot detect $B$'s attempt to front running, but   $B$ observes $A$'s commit message. In Section \ref{sec:hiding commitment}, we introduce a modified protocol allowing $A$ to hide her commit message.

Finally, we simplify the problem slightly by assuming that there is no state of the world $s$ such that $P_A(s)\in [c, c+\frac{c}{\beta}]$. Hence, if $A$ expects no front running, then the introduction of the protocol does not change the states of the world for which $A$ wants to send a message. This assumption is useful because it rules out situations in which $B$'s inference from observing that $A$ sent or not a message changes drastically with the introduction of the protocol.\footnote{An alternative that is equivalent for our purposes but more cumbersome is to assume that $s$ such that $P_A(s)\in [c, c+\frac{c}{\beta}]$ exist but are not very important from $B$'s viewpoint, in the sense that  $$\mbox{pr}\left\lbrace P_A(s)\in [c, c+\frac{c}{\beta}] \right\rbrace  E_s\left[P_B(\tilde \sigma_B, \sigma_A(s), s)| P_A(s)\in [c, c+\frac{c}{\beta}] \right]$$ is sufficiently small $\forall \tilde \sigma_B \in \Sigma$.  
}

\subsection{Equilibrium}

The first, rather immediate, result is that there is no equilibrium in which $B$ sends the same commit message as $A$. To see this, suppose that player $A$ sends the commit message    $S(addr,\sigma_A)$ and player $B$  sends the same commit message. If in the next period $B$ sends the message $reveal_B=\sigma_A$, then the SC will consider $B$'s reveal message as invalid because sent from an address  different from $addr$. It is also easy to see that there is no equilibrium in which $A$ commits and then does not reveal because $A$ can do better by not committing at all. The next lemma summarizes these observations.
\begin{lemma}[No cloning in equilibrium]
There is no equilibrium in which $\sigma_{B,1}=\sigma_{A,1}.$ There is also no equilibrium in which $A$ sends the commit message but not the reveal message.
\end{lemma}

In equilibrium, therefore, if $B$  wants to attack, he needs to craft a commit message while being  uninformed about the state of the world and $A$'s message. However, $B$ anticipates that he   will observe $A$'s message and, at that point, will decide whether or not to send the message he initially committed. Therefore, the protocol severely limits but does not fully eliminate $B$'s ability to act upon his observation of $A$'s message. Hence, it is possible that \eqref{eq: assumption} holds and, despite this, $B$ can profitably attack. 

Formally, suppose there is no front running in equilibrium so that $A$'s optimal message is again $\sigma^{**}_A(s)$ (see Equation \ref{eq: optimal no front running}).  Suppose $\sigma^{**}_A(s)\neq \emptyset$ (so that $A$ sent the commit message),  $B$ committed message   $\sigma_B$ and then observed $A$'s reveal message. In this case, $B$'s expected payoff from front-running is \[
q \cdot \tilde P_B(\sigma_B, \sigma^{**}_A(s), s) -f.
\]
Hence, $B$ will  try to front run if and only if $q \cdot \tilde P (\sigma_B, \sigma^{**}_A(s), s) > f$. 

In the commitment phase, $B$'s choice of what message to commit is made in anticipation that he will decide to front run after observing $A$'s reveal message. His expected future payoff is, therefore:
\[
\pi \equiv \mbox{max}_{\sigma_B\in \Sigma} E_s \left[\max \{ q \cdot \tilde P_B(\sigma_B, \sigma^{**}_A(s), s) -f,0\}| \sigma^{**}_A(s) \neq \emptyset \right],
\]
where the expectation is conditional on the state of the world being such that $A$ sends a commit message.
Hence, if $A$ sends a commit message and $B$ tries to front run, $B$'s expected payoff is $\beta \pi -c$. We therefore have the following proposition:\footnote{The existence of the equilibrium follows from the fact that the players' strategy space is finite, as noted already in \cite{nash1950equilibrium}. }
\begin{proposition}
    If $ \pi\leq \frac{c}{ \beta  }$ (i.e., ``guessing is hard for $B$''), then there is no front-running in equilibrium. If instead $ \pi > \frac{c}{ \beta  }$ (i.e., ``guessing is easy for $B$''), front running occurs with strictly positive probability in equilibrium.
\end{proposition}
\noindent Note that in case ``guessing is easy for $B$'', there could be a pure strategy equilibrium in which $B$ commits with probability 1 whenever $A$ commits, or a mixed strategy equilibrium in which $B$ commits with some probability. In either cases, after committing, $B$ attempts to front run $A$ or not depending on $A$'s reveal message.

It is easy to check that in the ``guessing is hard for $B$'' case, $A$'s equilibrium payoff is 
\[
 \max\left\lbrace -c + \beta (P_A(s)-c), 0 \right\rbrace  
\]
Therefore, the protocol generates both costs and benefits to player $A$. The main benefit is that the protocol reduces or eliminates front running. The  costs are two: one additional message is required, and the payoff is earned with a one-block delay (and hence is discounted by the parameter $\beta$). 



\section{Discussion}

\subsection{Severity of attacks}
The value of $\pi$  measures how easy it is for $B$ to guess. It is therefore   the inverse of the measure of severity of the attack discussed in the introduction: if it is difficult for $B$ to guess, it is because $B$ has very little prior information and, in the benchmark case, he relies heavily on observing $A$'s message, while the opposite is true when it is easy for $B$ to guess. Therefore, we can say that  the protocol is most effective at preventing the most severe front-running attacks.

\subsection{Attack vs legitimate competition}

We now compare the equilibrium  when $B$ is an attacker (i.e., condition \ref{eq: assumption} holds) with that when $B$ is a legitimate competitor (i.e., condition \ref{eq: assumption} is violated). To do so, we introduce the following condition
\begin{equation}\label{eq: assumption-strong}
q \cdot E_s[\tilde P_B(\tilde \sigma_B, \sigma^{**}_A(s), s)] \leq f +\frac{c}{\beta} ~~~\forall \tilde \sigma_B \in \Sigma, 
\end{equation}
which is akin to condition  \eqref{eq: assumption}, but where the cost of sending a message is now the cost of participating in the commit-reveal protocol.

Suppose first that the above condition is violated, that is, there exists a $\sigma^{**}_B \in \Sigma$ such that 
\[
-c+ \beta(q\cdot  E_s[\tilde P_B(\tilde \sigma^{**}_B, \sigma^{**}_A(s), s)] -f)> 0 
\]
This immediately implies that \eqref{eq: assumption} is violated and $B$ is a legitimate competitor. At the same time, we have
\begin{align*}
    -c\, + &\, \beta (q\cdot  E_s[\tilde P_B( \sigma^{**}_B, \sigma^{**}_A(s), s)] -f) < \\
           & \mbox{Pr}\left\lbrace \sigma^{**}_A(s) \neq \emptyset \right\rbrace  \cdot \left( -c + \beta E_s \left[\max \{ q \cdot \tilde P_B(\sigma^{**}_B, \sigma^{**}_A(s), s) -f,0\}| \sigma^{**}_A(s) \neq \emptyset \right] \right)  < \\
           & \mbox{Pr}\left\lbrace \sigma^{**}_A(s) \neq \emptyset \right\rbrace  \cdot \left( -c + \beta \pi \right)
\end{align*}
which implies that we are in the ``guessing is easy'' case.  Hence, if \eqref{eq: assumption-strong} is violated, then $B$ is a legitimate competitor and there is ``front running'' in equilibrium.


These derivations show that, modulo the fact that sending messages is more expensive with the protocol (i.e., the right-hand side of \ref{eq: assumption} is different from the right-hand side of \ref{eq: assumption-strong}), the protocol does not impede legitimate competition: both players  commit their messages and then compete with each other to have their reveal message included first in the following block. At the same time, attacks are more costly because an attacker is forced to make a costly guess. Hence, the protocol discourages front-running attacks (especially the most severe) while competition among honest players is preserved but postponed by one period.

\subsection{Additional messages}

In the above analysis, we restricted the players' action space to a single message per player in the commit period.  
However, if we relax this assumption, $B$ may want to send multiple commit messages.  For a given number of commit messages $k$, $B$ will choose the $k$ messages that jointly generate the largest expected payoffs $\pi(k)$, which can be defined similarly to the earlier case. The choice of how many messages to send is  also quite straightforward: it is the $k^*$ such that $\pi(k^*+1)-\pi(k^*)<c$. There is a ``guessing is hard'' case identical to the one discussed earlier. There is also a ``guessing is easy'' case, which is more convoluted  because the number of messages committed by $B$ may be greater than 1. However, the intuition is largely unchanged from the simple case.

\subsection{Pre-commitments} 
Another restriction we imposed is that the protocol starts when player $A$ learns the state of the world. It is, however, possible that $A$ may want to \textit{pre-commit}, that is, commit a message before learning the state of the world, in the hope that the committed message can be used immediately when the state of the world is revealed. The important observation is that $A$ can pre-commit and then decide to restart the protocol by committing a second message upon learning the state of the world. This complicates $B$'s inference problem because whatever message he commits may be wasted in the future. Again, the basic insight from the simple model above continues to hold, but guessing is  harder for player $B$.

\subsection{Multiple attackers}\label{sec: multiple attackers}

An interesting implication of our protocol is that it may reduce or eliminate competition between attackers, therefore benefiting the attackers as well as the honest player. To see this, assume that there are two attackers, $B_1$ and $B_2$. When sending a transaction, each $B_i$ chooses how much money to spend $f_i \geq 0$, which is now a continuous choice variable. The attackers' choices are made simultaneously and independently from each other. 

To remain as close as possible to the case with a single attacker (and leverage the results already derived),  we can think of competition between the two attackers and the honest player as happening in two steps. First, the attacker that spends the most wins the right to attack the honest player. Then, similarly to the single-attacker case, the winner attempts to front-run the honest player and, if successful, earns $\tilde P_B(\tilde \sigma_B, \tilde \sigma_A, s)$ (same for both attackers).\footnote{All our results are robust to other ways to model competition. The reason is that our results rely on  the full dissipation of rents: in the equilibrium of the contest game, the weakest attacker expects to earn zero. This result holds in a large class of contest models.} Mathematically, the probability that the transaction sent by player $B_i$ is included in the block before that of $B_{-i}$ and player $A$ is:
\[
\begin{cases}
 \gamma_i q (f_i)   &\mbox{ if} f_i > f_{-i}\\
 0 &\mbox{ if} f_i < f_{-i}
\end{cases} 
\]
where the function $q():\mathbf{R}^+ \rightarrow [0,1]$ is strictly increasing, strictly concave, and differentiable. The parameter $\gamma_i>0$ for $i\in\{1,2\}$ measures the strength of each attacker. Without loss of generality, we assume that the attacker number 1 is stronger, and hence $\gamma_1 \geq \gamma_2$. A tie-breaking rule determines what happens when $f_i=f_{-i}$, but the nature of such a rule is not important for our analysis. 

\paragraph{No commitment needed.} We start by deriving what happens with multiple attackers when players can send their messages directly to the smart contract. After observing the victim's message and learning the state of the world, attacker $B_i$'s payoff as a function of $f_i, f_{-i}$ is
\[
\begin{cases}
 P_B(s) \gamma_i q (f_i) -f_i  &\mbox{ if }  f_i > f_{-i}\\
 -f_i &\mbox{ if } f_i < f_{-i}
\end{cases} 
\]
Formally, the attackers are engaged in an asymmetric contest with productive effort, as studied in \cite{siegel2014contests}. We can therefore use its Theorem 1 to characterize the equilibrium of the game.

Define 
\[
\underline f_i \equiv \mbox{argmax}_{f_i} \left\lbrace P_B(s) \gamma_i q (f_i) -f_i \right\rbrace 
\]
as the optimal expenditure by attacker $i$ whenever attacker $-i$ is absent and 
\[
\overline f_i \equiv f_i: P_B(s) \gamma_i q(f_i) =f_i
\]
as the expenditure level at which attacker $i$'s payoff is zero  whenever attacker $-i$ is absent. Clearly, whenever $\underline f_1 \geq  \overline f_2 $, then there is a unique equilibrium in pure strategy, in which attacker $B_1$ sets $f^*_1=\underline f_1$ and attacker 2 does not do anything. This situation is therefore identical to the single-attacker case discussed earlier.

If instead $\underline f_1 < \overline f_2 $, according to Theorem 1 in \cite{siegel2014contests}, there are multiple mixed-strategy equilibria. However, in every equilibrium of the game attacker 1's payoff is 
\[
P_B(s) \gamma_1 q (\overline f_2)-\overline f_2.
\]
That is, the strong attacker's payoff is equal to the payoff he would achieve if he'd set his expenditure equal to the follower's largest possible expenditure.\footnote{This result is also in \cite{siegel2009all}, in which however only non-productive effort is considered.  \cite{siegel2014contests} extends these results to cases in which, over some range, the ``prize'' to be won by a player may be increasing in this player's effort.} The utility of the other attackers is zero.

To summarize, relative to the single-attacker case, if there are two attackers who are sufficiently similar then in equilibrium they will randomize their level of spending. In expectation, the weaker attacker earns zero. The stronger attacker earns a positive amount, which is however lower than if he was the unique attacker. Competition, therefore, hurts both attackers because they overspend (relative to the single-attacker case). 

\paragraph{Commit-reveal protocol.} Consider the commit-reveal protocol. At the beginning of the commit period player $A$ sends the commit message, which is observed by the attackers who simultaneously choose their commit message. For simplicity, we assume that the attackers can observe each other's commit messages after they are set.\footnote{The attackers send their commit messages simultaneously, so whether these messages are observable does not affect the commitment stage of the game. In the subsequent reveal stage, if an attacker did not observe the other attacker's commit message, he will nonetheless detect the opponent's attempt to front run. At that point, he will increase his level of spending. The outcome is identical to the case in which, at the beginning of the reveal stage,  the attacker knows  that the other attacker committed and will therefore attack. } We solve the game backward, starting from the reveal phase.

If only one attacker $B_i$ committed, then the problem is quite simple: the single attacker $i$ earns\footnote{Remember that the attacker has the same payoff function and information. Hence, in the commit period, if they commit they will both commit $\hat \sigma_B$.}
\[
V(\gamma_i) \equiv \max_{f_i} \left\lbrace  \tilde P_B (\tilde \sigma_B, s) \gamma_i q(f_i) -f_i \right\rbrace
\]
If instead both attackers committed and they are sufficiently similar, then the logic discussed in the previous section continues to apply:  both attackers overspend, and the weaker attacker expects to earn zero while the stronger attacker expects to earn $\underline V (\gamma_1) < V(\gamma_1)$.\footnote{The  meaning of ``the attackers being sufficiently similar'' and the expected payoff of player 1 can be precisely derived following the same steps illustrated in the previous paragraph. But their precise expressions are not important in what follows.}   

Given this, we can derive the equilibrium in the commitment phase. The main result is that there is no equilibrium in which both players commit with probability 1. The reason is that if such an equilibrium existed, the weak attacker would earn zero after committing. Because commitment messages are costly,  the weak attacker is better off not committing.

It follows that the equilibria of the game are
\begin{itemize}
    \item if either $\beta \underline V(\gamma_1) >c$, or  $\beta V(\gamma_1) >c> \beta \underline V(\gamma_1) $ and $c>\beta V(\gamma_2)$, then there is a unique equilibrium in pure strategy in which only the strong attacker (attacker 1) commits.
     \item if $\beta V(\gamma_1) >c> \beta \underline V(\gamma_1) $ and $\beta V(\gamma_2) >c$, then there are two pure strategy Nash equilibria, each corresponding to only one attacker sending the commit message. There is also a mixed strategy equilibrium, in which attacker 1 commits with probability $\alpha_1$ and attacker 2 commits with probability $\alpha_2$. These probabilities are such that each attacker is indifferent between committing or not, that is $\alpha_1 V(\gamma_2)=c$ and $\alpha_2 V(\gamma_1) + (1-\alpha_2) \underline V (\gamma_1)=c$. In this equilibrium, there is a probability $\alpha_1 \alpha_2$ that both attackers commit, a probability $(1-\alpha_1)(1- \alpha_2)$ that no attackers commit, and the remaining probability that a single attacker commits. 
     \item otherwise, no attacker commits, and front running is prevented. 
\end{itemize}

The protocol, therefore, decreases the level of competition among attackers. This, in turn, increases $A$'s payoff as it reduces the aggregate expenditure  by the attackers.

\section{Extensions to the protocol}
\subsection{Hiding commitments}\label{sec:hiding commitment}


Here we propose a version of the protocol that allows hiding the commit message. The modified protocol exploits the fact that player $A$ can send commit and reveal messages from different addresses, provided that the commit message includes the address that $A$ will use in the following period to send the reveal message.

To study how the possibility of hiding the commit messages affects the equilibrium of the game, here we assume that player $A$ observes the state of the world only with some probability, in which case she may decide to send her message. If instead, she does not observe the state of the world, then she takes no action. We also replicate the game $n$ times: there are now $n$ identical $A$ players who, with some probability may want to interact with one of $n$ smart contracts. 

\begin{figure}[ht]
    \centering
    \includegraphics[width=15cm,trim={0 0cm 0 0cm}]{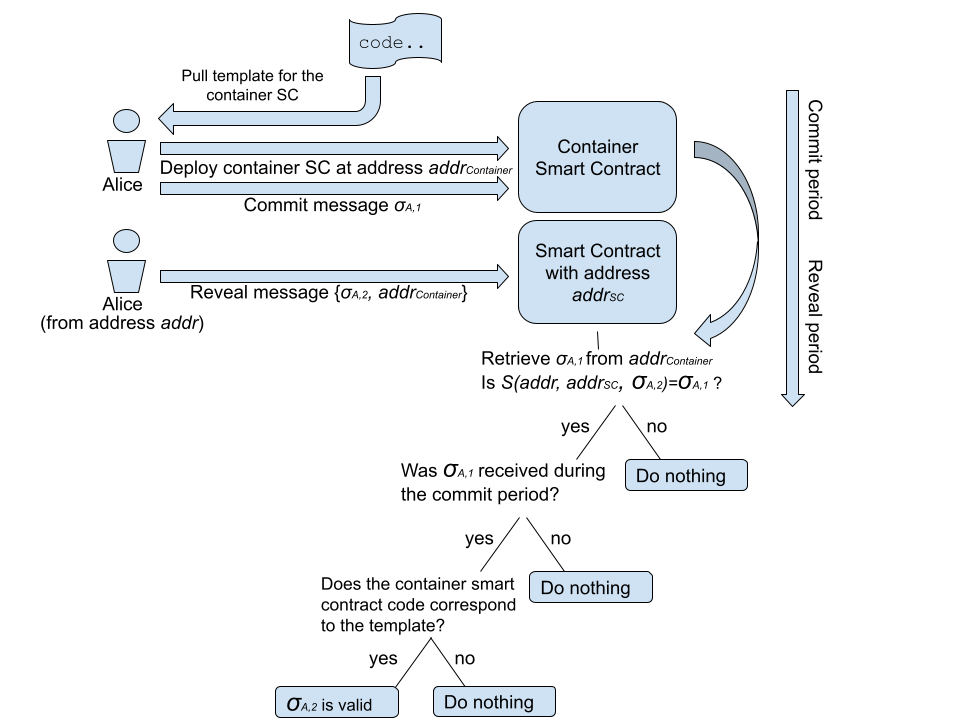}
    \caption{The commit-reveal protocol with obfuscation}
    \label{fig:complex}
\end{figure}

These modifications are irrelevant in the protocol that we introduced earlier, because, in each replica game, $B$ can send his commit message  after observing $A$'s commit  message.  These modifications however become relevant if we modify the protocol  so that both the sender and the receiver of the commitment are obfuscated. More precisely, the modified protocol is now (see also Figure \ref{fig:complex}):
\begin{itemize}
    \item there is a pre-existing template code for a \textit{container smart contract}. This code is such that when the container smart contract receives a commit message, it time-stamps it with the current block number.
    \item to commit, the honest player generates a brand-new address and uses it to send a transaction in which, first, a container smart contract is created using the template, and then the commit message is sent to the newly-created container smart contract.\footnote{The brand new address needs to be founded with some $ETH$ before it can send messages. We note that this could be done via a centralized exchange, therefore hiding the identity of the creator of the new address from the attacker.} The commit message is now $S(addr,addr_{SC},\sigma_A)$, where $addr_{SC}$ is the address of the target smart contract.
    \item to reveal, the honest player sends  the message $\sigma_A$  to the target smart contract, together with a pointer to the container smart contract in which the commit message is stored.
    \item the target smart contract considers the message as valid if all these conditions are satisfied
    \begin{enumerate}
        \item the commit message should be $S(addr,addr_{SC},\sigma_A)$, where $addr$ is the address from which the reveal message was sent. 
        \item  the timestamp associated with the commit message is lower than the current block number. This step makes sure that the commit message was sent before the reveal message.
        \item the code of the container smart contract is identical to the template smart contract.\footnote{In Ethereum, a smart contract code is accessible by other smart contracts. For example, the  expression $type(SC).creationCode$ returns the creation bytecode of smart contract $SC$ (see \url{https://docs.soliditylang.org/en/latest/units-and-global-variables.html\#type-information}). If the template storage contract specifies that the contract is immutable, such bytecode will be constant and cannot be changed. 
        } 
        \andrea{with respect to the footnote, this paper claims that no smart contract is immutable anymore on Ethereum \url{https://informationsecurity.uibk.ac.at/pdfs/FB-Ethereum-Create2.pdf}  replacing the code of a SC cannot be done in a single transaction (you need to retire the code first, and then deploy the new code), but it can be done within the same block. Or, at least, this is what I gather from their footnote 7}
    \end{enumerate}
\end{itemize}
\noindent The fact that the commit message includes $addr_{SC}$ forces an attacker to send a different commit message for each honest user/smart contract pair he may attack. The very last step is necessary to prevent an attack in which, after observing the reveal message, an attacker sends a single transaction that (i) creates a container smart contract, (ii) stores the commitment there together with a fake time stamp  and (iii) sends the reveal message. 

An outside observer can infer that someone created a container smart contract using the template and committed something, but does not know who committed nor the target smart contract that will receive the reveal message. Call $\tau$ the ratio between the observed container smart contracts created and $n$. 
The same logic discussed above implies that if $\tau  \cdot \pi \leq c/\beta$, then it is too costly for $B$ to attack a given $A$ player: guessing is too hard for $B$ and front-running is prevented. Hence, if the probability that a given honest user sends a message to a given smart contract is sufficiently low (so that the realized $\tau$ is also low), then the protocol eliminates all front-running attacks.

An important observation is that the above scheme is effective in hiding the target smart contract if and only if multiple target smart contracts share the same template for the container smart contract. In the extreme case in which each target smart contract has its own template, then the identity of the user remains hidden but the target smart contract that will receive the reveal can be inferred. At the other extreme, the highest level of obfuscation is achieved when all smart contracts use the same template. Different smart contracts could also coordinate by creating a single ``official'' container smart contract that receives all commitments. Again, an outside observer can infer that a user sent a commitment to the container smart contract, without knowing who is the user and what is the target smart contract. At the same time, users do not need to create the container smart contract each time, leading to significant savings in gas.  How to achieve this coordination among smart contracts is not part of the model.

\andrea{Here perhaps we may include a discussion on time-locks: preventing $B$ from acting on $A$'s reveal message}

\subsection{Partial implementation.}\label{sec: partial implementation}
It is possible to implement the protocol only  for a subset of  messages. That is, there is a set of messages $M\subset \Sigma$  such that any message $\sigma\in M$ is considered valid by the SC only if the commit-reveal protocol described above is followed. All other messages $\sigma \not\in M$ are considered valid by the SC as soon as they are received. 

Suppose that $A$ wants to send message $\sigma_A$ and $B$ wants to front run with message $\sigma_B$. There are four possible cases:
\begin{enumerate}
\item $\sigma_A, \sigma_B \in M$, which means that we are in the commit-reveal case discussed earlier.
\item $\sigma_A, \sigma_B \not\in M$, which means that we are in the benchmark case discussed earlier.
\item $\sigma_A \not\in M$ but $\sigma_B \in M$, which means that $A$ can send her message directly but $B$ needs to commit his message and then reveal it. In this case, front running is prevented at no cost for $A$.
\item $\sigma_A \in M$ but $\sigma_B \not \in M$, which means that $A$ needs to send two messages (commit and reveal), and wait one period, for in the end a probability of being front-run which is the same as in the benchmark case. In this case, the protocol imposes extra costs on $A$ for no benefit.
\end{enumerate}
The specific design of $M$ depends on the situation and will balance the possible costs and benefits to player $A$. With this respect, an important observation is that the choice of  $M$ determines  $\pi$. So, for example, for given   $\pi$, it would seem beneficial not to use the protocol in states of the world in which player $A$ does not expect an attack. But this may not be optimal, because states of the world where $A$ does not expect to be attacked are precisely the ones in which the attackers' payoff is low. Hence, by applying the protocol also in these states of the world, $\pi$ decrease, and with it the probability of a front-running attack. 

\subsection{Specifying commit and reveal periods}\label{sec: waiting in the mempool}
    
Our model assumes that both commit and reveal messages are included in a block immediately after being sent.  In practice, however, messages may remain in the mempool for some time before being included in a block.\footnote{We treat this possibility as a random event, not something that an attacker could manipulate. The reason is that purposefully censoring  a transaction requires a large number of miners/validators to collude, which is prevented by the consensus protocol.   } This possibility is not an issue with respect to the commit message, because the honest player can simply wait until this message is included in a block before sending the reveal message. It is however an issue with respect to the reveal message, because an attacker may be able to observe the victim's reveal message, send a commit message, have it included in a block, then send a reveal message, and have it included in a block before the honest user's reveal message.

To start, note that the possibility that messages stay in the mempool is a concern also in the benchmark case (i.e., the standard way in which Ethereum operates), possibly even more than in our protocol  because an attacker needs to send just one message during the period in which the honest player's message stays in the mempool (vs two in our protocol). It is also a concern that is greatly reduced by the introduction of the base fee: a fee that should guarantee the rapid inclusion of a transaction in a bloc (see footnote \ref{footnote: EIP1559}).

For our purposes, it is interesting to note that  our protocol can reduce or eliminate this concern by appropriately specifying commit and reveal periods, that can be thought of as sets of blocks. The $SC$ will then consider a reveal message as valid only if received during a block belonging to the reveal period, and only if its commit message was received (either directly by $SC$ or via the container smart contract) during a block belonging to the commit period.

For example, a specific application may have a natural deadline, such as  a competition rewarding whoever can provide the  correct answer to a question within a specific time period. In these situations, it seems natural to specify the commit period as all blocks up until the deadline and the reveal period as all blocks after the deadline, therefore eliminating the risk that an attacker commits after having observed the reveal message. In other situations, it may be possible to alternate between commit periods and reveal periods. In this case, the above attack is possible only if the reveal message remains in the mempool for the entire length of the reveal period---a probability that drops to zero rapidly with the length of this period. Of course, this modification has a cost because it increases $A$'s waiting time (i.e., the time between $A$  learning the state of the world and deciding to send her message and the time he receives her reward).

Finally, it is also possible that the commit-reveal protocol is required only in some periods. For example, during the ``commit'' period users could either commit or send a message directly to the smart contract without any commitment, which would be considered valid. In the reveal period, only reveal messages that were committed during the commit period are considered valid.  The honest player can choose to send a given transaction in a ``slow but safe'' mode, or a ``fast but risky'' mode. In the slow but safe mode, the user sends her commitment during the ``commit'' period and the reveal in the ``reveal'' period, therefore preventing an attacker from sending both commit and reveal messages after observing the honest player's reveal message. In the fast but risky mode, a user sends a direct message to the smart contract during the commit period. Doing so exposes the honest player to the risk of being front-run but may nonetheless be optimal if the honest player is particularly impatient.

\section{Conclusion}

We conclude by discussing some limitations to our protocol that require further study.


Our commit-reveal protocol may  impede the possibility of calling different smart contracts within the same transaction (usually referred to as smart-contract \textit{composablity}). In principle, composability is still possible by first committing the different messages to the various smart contracts. However, a problem arises when these smart contracts have different commit-reveal periods (see Section \ref{sec: waiting in the mempool}). Although different commit messages may be sent in different periods depending on the commitment window of each smart contract, to maintain composability the reveal messages must be sent within the same transaction during the reveal window of all smart contracts. If such a window does not exist, then composability is not possible. If it  exists, then it is possible, but exploiting it may impose large delays to the execution of the transaction. Studying further how to mitigate this problem is  left for future work. Here we note that composability is preserved  if the commit-reveal protocol is required only in some periods (as discussed in the last paragraph of Section \ref{sec: waiting in the mempool}), chosen in a coordinated way among all smart contracts. 

Our analysis assumes that the smart contract does not have an explicit mechanism to resolve competing claims to an object and therefore does not apply to, for example, a smart contract running an on-chain auction. 
Applying our protocol to such a smart contract may lead to unintended consequences because the players (honest or not) may fail to reveal after committing, perhaps because they realize they would lose the auction. This is problematic in many cases. For example,  failure to reveal in a second-price auction may decrease the revenues raised.


Our protocol is also not effective against a type  of front-running attack called \textit{suppression attacks }in which an attacker prevents the victim's transaction from being included in a block by front-running it with a series of spam transactions (see \cite{eskandari2019sok}). The reason is that, in these attacks, the content of the victim's transaction is irrelevant to the attacker. However, these types of attacks are rare and specific to certain applications. For example,  \cite{eskandari2019sok} document only one of them in the context of a gambling smart contract.

Finally, our analysis was purposefully general. However, we recognize that the details of the implementation of the protocol in a specific context are crucial as they determine whether its benefits outweigh its cost. Considering a specific application and deriving which messages should require a prior commitment (see Section \ref{sec: multiple attackers}) and how to specify commit and reveal periods (see Section \ref{sec: waiting in the mempool}) is left for future work.

\bibliography{bib}{}
\bibliographystyle{chicago}

\pagebreak
\appendix

\end{document}